\documentclass[a4paper,prstab,twocolumn,nofootinbib,superscriptaddress,floatfix,amsmath,amsfonts,amssymb,nomerge]{revtex4-1}

\usepackage[utf8]{inputenc}
\usepackage[english]{babel}
\usepackage{graphicx}
\usepackage{epsfig}
\usepackage{hyperref}
\usepackage{wasysym}
\usepackage[percent]{overpic}
\usepackage{xcolor}

\begin{document}

\title{Influence of proton bunch parameters on a proton-driven plasma wakefield acceleration experiment}

\author{Mariana Moreira}
\email{mariana.moreira@cern.ch}
\affiliation{GoLP/Instituto de Plasmas e Fus\~ao Nuclear, Instituto Superior T\'ecnico, Universidade de Lisboa, 1049-001 Lisboa, Portugal}
\affiliation{CERN, CH-1211 Geneva, Switzerland}

\author{Jorge Vieira}

\affiliation{GoLP/Instituto de Plasmas e Fus\~ao Nuclear, Instituto Superior T\'ecnico, Universidade de Lisboa, 1049-001 Lisboa, Portugal}

\author{Patric Muggli}

\affiliation{Max Planck Institute for Physics, D-80805 Munich, Germany}
\affiliation{CERN, CH-1211 Geneva, Switzerland}

\begin{abstract}

We use particle-in-cell (PIC) simulations to study the effects of variations of the incoming 400~GeV proton bunch parameters on the amplitude and phase of the wakefields resulting from a seeded self-modulation (SSM) process.
We find that these effects are largest during the growth of the SSM, i.e. over the first five to six meters of plasma with an electron density of $7 \times10^{14}$~cm$^{-3}$.
However, for variations of any single parameter by $\pm$5\%, effects after the SSM saturation point are small.
In particular, the phase variations correspond to much less than a quarter wakefield period, making deterministic injection of electrons (or positrons) into the accelerating and focusing phase of the wakefields in principle possible.
We use the wakefields from the simulations and a simple test electron model to estimate the same effects on the maximum final energies of electrons injected along the plasma, which are found to be below the initial variations of $\pm$5\%. This analysis includes the dephasing of the electrons with respect to the wakefields that is expected during the growth of the SSM. Based on a PIC simulation, we also determine the injection position along the bunch and along the plasma leading to the largest energy gain.
For the parameters taken here (ratio of peak beam density to plasma density $n_{b0}/n_0 \approx 0.003$), we find that the optimum position along the proton bunch is at $\xi \approx -1.5 \; \sigma_{zb}$, and that the optimal range for injection along the plasma (for a highest final energy of $\sim$1.6~GeV after 10~m) is 5--6~m.

\end{abstract}

\maketitle

\section{Introduction}

The AWAKE experiment intends to demonstrate the concept of proton-driven plasma wakefield acceleration using 400\,GeV proton bunches supplied by the Super Proton Synchrotron (SPS) at CERN to accelerate externally injected electrons~\cite{awake}.
The concept underlying AWAKE is one of several that have been proposed for plasma-based acceleration, which could pave the way towards higher collision energies than what conventional accelerator technology can provide.
An estimate for the maximum acceleration gradient supported by plasma is given by the cold non-relativistic wavebreaking field~\cite{akhiezer,dawson}
\begin{equation}
E_0 = \frac{m_e c \, \omega_{pe}}{e} \, \approx \, 96 \sqrt{n_0 \left[ \mathrm{cm}^{-3} \right]} \, \left[ \mathrm{V/m} \right] \; ,
\end{equation}
where $c$ is the speed of light, $m_e$ is the electron mass, $e$ is the elementary charge, $n_0$ is the plasma electron density, and $\omega_{pe} = \sqrt{e^2 n_0 / \varepsilon_0 \, m_e}$ is the electron plasma frequency and $\varepsilon_0$ is the vacuum permittivity.
The plasma density used in AWAKE, for example, of the order of $10^{14} \; \mathrm{cm}^{-3}$, yields $E_0 \approx 1 \; \mathrm{GV/m}$, which is approximately ten times larger than what is feasible with RF cavities at the moment~\cite{rfcavities}.
For higher plasma densities ($10^{18} \; \mathrm{cm}^{-3}$), however, acceleration gradients of the order of $100 \; \mathrm{GV/m}$ could be reached.

Plasma-based acceleration can be accomplished using either a laser pulse or a particle bunch as a driver.
AWAKE is an instance of the latter case, which is also known as plasma wakefield acceleration~\cite{pwfa} (PWFA).
As a particle bunch propagates in plasma, the fields caused by
its space charge disturb the light plasma electrons, while the more massive plasma ions can be assumed to remain immobile (at the 1/$\omega_{pe}$ time scale) as long as the ion to electron mass ratio is sufficiently high~\cite{vieiraiona,vieiraionb}.

The displaced plasma electrons in the wake of the particle driver oscillate at the plasma frequency $\omega_{pe}$, and this density oscillation is in turn associated with transverse and longitudinal fields, the wakefields.
The wavelength of the resulting plasma wave (or wake) is thus related to $\omega_{pe}$ and is called the plasma wavelength: $\lambda_{pe} = 2 \pi v_b / \omega_{pe}$, where $v_b\simeq c$ is the proton bunch velocity.

When the drive bunch is short, i.e. with a typical length $L \apprle \lambda_{pe}$, the wake travels with the speed of the driver.
A charged particle can then be trapped and accelerated if it is injected with roughly the same speed as the plasma wake in a region of the wakefields that is longitudinally accelerating and transversely focusing.
In the linear regime, where the beam density of the drive bunch $n_b$ is much smaller than the plasma density ($n_b \ll n_0$), the transverse and longitudinal components of the wakefields are harmonic and phase-shifted by a fourth of a period with respect to each other, as expressed by a unique relationship between both components known as the Panofsky-Wenzel theorem~\cite{PWtheo}.
This means that each ideal region for acceleration, where the fields are both accelerating and focusing, is $\lambda_{pe}/4$ long.

In order to drive the wakefields effectively, the length of the driver should be of the order of $\lambda_{pe}$.
This is not the case in AWAKE, where the bunches delivered by the SPS are considerably longer (6--12~cm) than the plasma wavelengths in the adjustable density range ($\sim$1--3~mm for (1--10) $\times 10^{14} \; \mathrm{cm}^{-3}$).
This causes the long proton bunch to undergo the self-modulation instability (SMI)~\cite{kumarsmi}, whereby the bunch is progressively modulated into a train of shorter bunches, with lengths and separation distances of the order of $\lambda_{pe}$, due to periodic transversely focusing and defocusing fields.
This instability eventually saturates and the initial proton bunch is self-consistently transformed into a bunch train, a format that can resonantly excite the wakefields.

The onset of an instability can either be due to noise or to a seed, i.e. a signal of higher amplitude than the noise level.
When the SMI starts from noise, both the phase of the wakefields along the bunch as well as their amplitude vary randomly from event to event and thus prevent reliable acceleration of injected particles.
In principle, seeding the instability is a means to fix the final phase and amplitude of the wakefields once the process has saturated.
The process is then called seeded self-modulation (SSM)~\cite{awake}.
Seeded self-modulation was recently demonstrated experimentally using a sharp ionisation front created by an optical laser within the long proton bunch~\cite{adliprl,turnerprl}.

It has been shown both theoretically~\cite{schroeder} and through numerical simulations~\cite{pukhov} that the phase velocity of the wakefields is smaller than that of the drive bunch during the growth of the SMI.
This limits the maximum energy gain since electrons can easily find themselves in the defocusing and decelerating phase of the wakefields and be lost.
External injection must therefore occur near or after saturation, when the wakefield phase velocity is very close to the driver velocity~\cite{pukhov}.
In addition, for this injection to succeed reliably, as is required for the application as a particle accelerator, the injected bunch must be deterministically placed in the accelerating and focusing phase of the wakefields, or within a range of $\lambda_{pe}/4$.
The wakefield phase at the point of injection along the proton bunch and along the plasma must therefore be reproducible to within a fraction of that range.
This must be true even in the presence of natural fluctuations of the drive bunch and of the accelerating structure, in this case the plasma.
It is therefore essential to study the effect of parameter variations on the wakefield characteristics.
Here we will assume that the plasma density and thus the frequency of the wakefields does not vary.
This is an assumption that is addressed in experiments by carefully controlling the plasma parameters~\cite{vaporsource}.

In this work we focus on the effects of bunch parameter and plasma radius fluctuations on the amplitude and phase of the wakefields after saturation of the SSM, where acceleration over a long distance can in principle start~\cite{savard}. We then use test electron calculations to infer the same effects on the energy of the accelerated electrons, and to study the optimal injection conditions that lead to the most acceleration.

The effects of initial bunch parameter variations are studied through numerical particle-in-cell (PIC) simulations in two-dimensional, axisymmetric cylindrical coordinates, performed with the code OSIRIS~\cite{osiris,osiris2}.
The values of a set of proton bunch parameters are varied independently and the respective simulations compared to a baseline simulation with parameters similar to those of AWAKE.
We note here that the hose instability, which can possibly compete with the SMI~\cite{hose}, is not described in 2D axisymmetric geometry.
We therefore assume in this work that the seed for the self-modulation process is large enough to prevent the growth of the hose instability~\cite{kumarsmi,vieirahose}.

\section{Simulation and parameters}

In the simulations used for this work, a moving window approximately 33~cm long and 1.6~mm high moves at $c$ with a proton bunch (moving at $\sim c$) as the latter propagates through 10~m of plasma.
The simulation box consists of a grid of 20063 cells in the longitudinal and 425 cells in the transverse direction, which corresponds to a resolution of roughly $17 \; \mathrm{\mu m}$ and $4 \; \mathrm{\mu m}$ (or 74 and 333 cells per $\lambda_{pe}$), respectively.
There are four particles per cell for each particle species (plasma electrons and beam protons) in the simulation.

The proton beam propagates with a Lorentz factor $\gamma_b \approx 480$ (corresponding to 450~GeV) with an energy spread of 0.035\% and a normalized emittance of 2.5~mm~mrad.
The profile of the proton bunch is implemented with a sharp cut, which represents the plasma creation by the co-propagating laser pulse, i.e. the relativistic ionization front, that seeds the SSM process in the experiment.
In these simulations the seeding of the self-modulation process is thus modeled by the sharp rising edge of the proton bunch.
The bunch density profile is given by
\begin{multline} \label{eq:nb}
n_b(\xi,r) = \\ \frac{ n_{b0} }{2} \left[ 1 + \cos \left( \sqrt{ \frac{\pi}{ 2 \sigma_{zb}^2 } } \left( \xi - \xi_s \right) \right) \right] e^{-r^2/\left( 2 \sigma_{rb}^2 \right) } \; ,
\end{multline}
for $\xi_0 \le \xi \le \xi_s$, where $\xi$ is the beam co-moving coordinate defined as $\xi = z - c t$, $n_{b0}$ is the peak bunch density, $\sigma_{zb}$ and $\sigma_{rb}$ are the RMS bunch length and width, respectively, $\xi_0$ is the position where the function crosses the $\xi$ axis (end of the bunch at $\xi_0=-\frac{\pi\sigma_{rb}}{2}$ for $\xi_s=0$), and $\xi_s$ is the seed position along the bunch.
The plasma fills the simulation window up to the ionization radius $r_p = 1.5\;\mathrm{mm}$.

The following parameters were used in the simulations: $n_0 = 7 \times 10^{14} \; \mathrm{cm^{-3}}$, $\sigma_{zb} = 12.6 \; \mathrm{cm}$, and $\sigma_{rb} = 200 \; \mathrm{\mu m}$.
The peak density in Eq.~\ref{eq:nb} is calculated according to
\begin{equation} \label{eq:nb0}
n_{b0} = \frac{N_b}{(2 \pi)^{3/2}\, \sigma_{rb}^2 \, \sigma_{zb} } \; ,
\end{equation}
giving $n_{b0} \approx 1.89 \times 10^{12} \; \mathrm{cm^{-3}} $ for the proton bunch population $N_b = 1.5 \times 10^{11}$ in the full bunch.

The following parameters were independently varied by $\pm 5 \%$: $\sigma_{zb}$, $\sigma_{rb}$, $N_b$ and $r_p$.
The RMS timing jitter of the proton bunch with respect to the ionizing laser pulse $\Delta t$ was also varied by $\pm 15 \; \mathrm{ps}$.
Note that $\Delta t$ is in practice a phase shift of the cosine in Eq.~\eqref{eq:nb} with respect to the center of the profile $\xi_s$, thus encompassing either more or less charge depending on whether the maximum of the cosine is moved to the right or left of $\xi_s$.

These parameters are taken as representative for the AWAKE experiment. However, we expect the conclusions presented here to be quite general.
In fact, we have confirmed that our conclusions will hold, by performing additional simulations with a new set of initial conditions (e.g. doubling the bunch charge).

\section{Properties of the wakefields}

\begin{figure}[t]
\includegraphics[width = \linewidth]{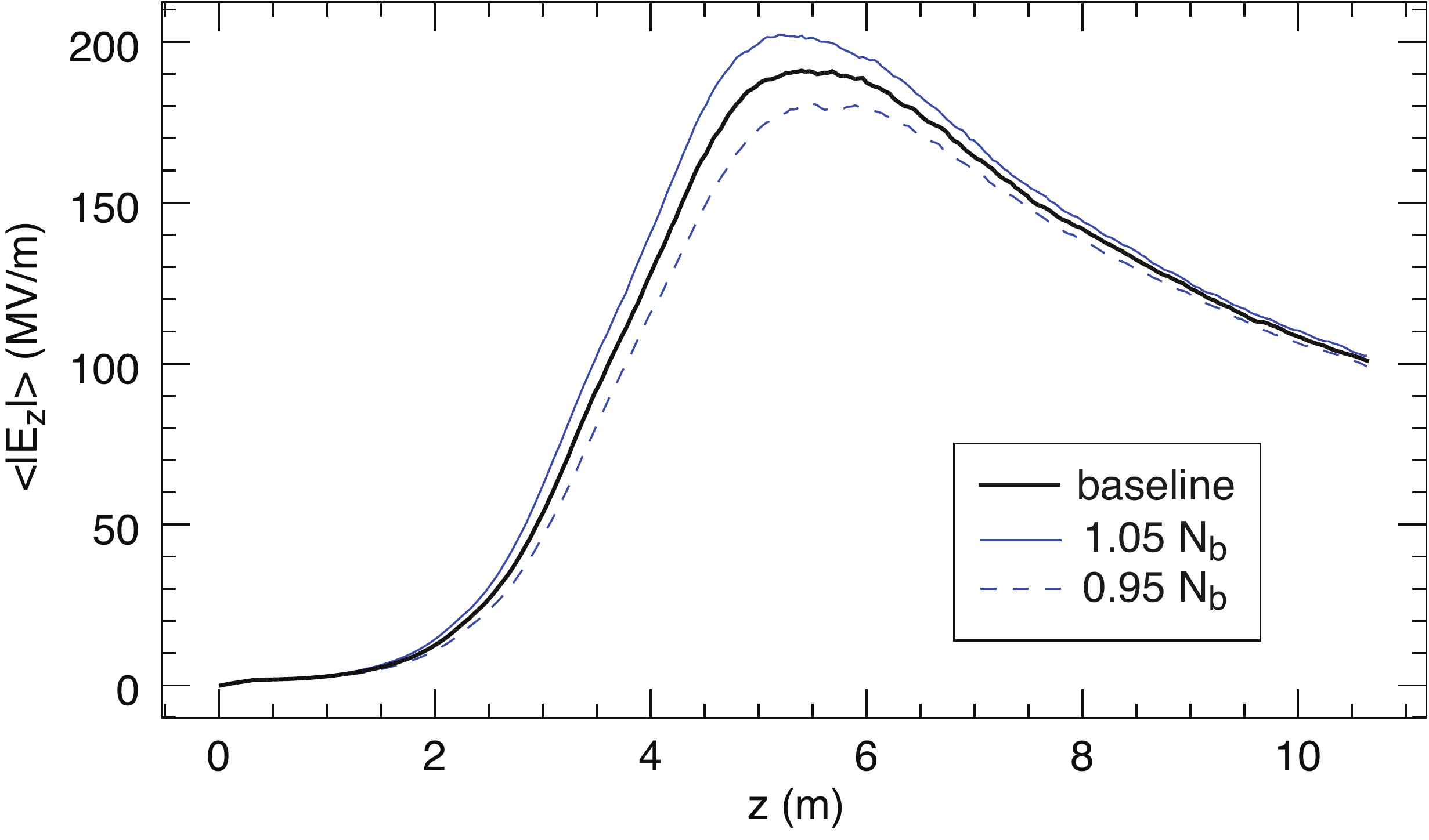}
\caption{\label{fig:e1ave} Average absolute value of $E_z$ for $0 < r < k_{pe}^{-1}$ as a function of the propagation distance $z$ for the baseline simulation and for the $\pm5\%$ variations in $N_b$.}
\end{figure}

A reliable plasma accelerator necessarily requires both amplitude and phase stability of the wakefields in the face of natural drive bunch parameter fluctuations.
Phase stability is especially critical since the accelerated electrons may otherwise slip into defocusing and decelerating regions of the wakefields and be lost before gaining a significant amount of energy.

Both the wakefield amplitude and the SSM growth rate depend on the bunch density.
Wakefields driven by each self-modulated microbunch can reach an amplitude of the order of $E_z=\frac{n_{b0}}{n_0}E_{0}$ (in the linear regime).
Therefore, at a given plasma density, variations of the wakefields with respect to bunch parameters are expected to follow dependencies similar to that of $n_{b0}\propto\frac{N_b}{\sigma_{zb}\sigma_{rb}^2}$ (see Eq.\ref{eq:nb0}).

The effects of the bunch parameter variations on the wakefield amplitude were characterized by comparing the average absolute value of the oscillating field $E_z$ along the propagation distance $z$ ($\left< |E_z| \right>$) for each parameter. The average $\left< \cdot \right>$ is computed from $E_z$ values in the simulation window at radii smaller than the plasma skin depth $k_{pe}^{-1} = c/\omega_{pe}$ ($k_{pe}^{-1} \approx 201 \; \mathrm{\mu m}$ for $n_0 = 7 \times 10^{14} \; \mathrm{cm^{-3}}$).
This limit corresponds to the radial extent beyond which the proton-driven plasma wakefields become negligible.

\begin{figure}[t]
\includegraphics[width = \linewidth]{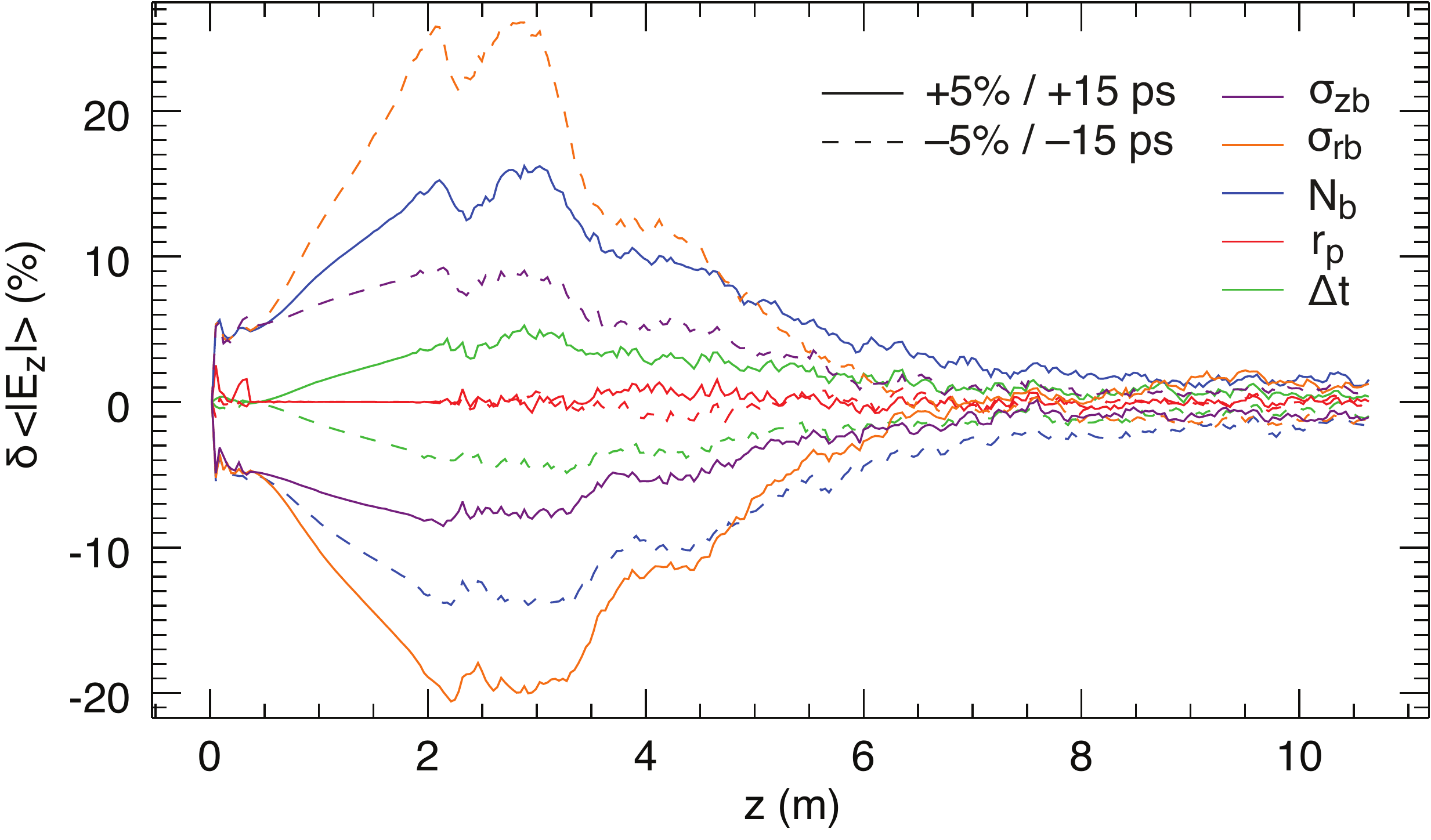}
\caption{\label{fig:e1averel} Relative deviation of the average absolute value of $E_z$ resulting from the parameter scans with respect to the baseline result (see Fig.~\ref{fig:e1ave}), where $\delta \left<|E_z|\right> = \left( \left<|E_z|\right> - \left<|E_{z, \mathrm{baseline}}|\right> \right) / \left<|E_{z, \mathrm{baseline}}|\right> $.}
\end{figure}

For example, the evolution of $\left< |E_z| \right>$ is shown in Fig.~\ref{fig:e1ave} for the baseline parameters and for variations in the bunch population $N_b$.
In the three cases the average fields grow rapidly until around $z = 4\; \mathrm{m}$, signifying the growth of the SSM, after which the SSM process saturates and the overall amplitudes of the wakefields gradually decrease.
We note here that this amplitude decay can in principle be avoided by using a small step in the plasma density early along the bunch propagation~\cite{lotovpop}.
As expected, more (less) bunch charge leads to a higher (lower) field amplitude.
These profiles are typical of all the simulations in this study.

The relative difference in $\left< |E_z| \right>$ with respect to the baseline simulation is shown in Fig.~\ref{fig:e1averel} for all the parameter variations.
In general, the effects of the parameter variations are maximum during the growth of the SSM ($z < 4 \; \mathrm{m}$), reaching a relative difference with respect to the baseline of approximately 26\% at $z \approx 2.8\; \mathrm{m}$ for $0.95 \; \sigma_{rb}$.
However, if electrons are injected only after the SSM process has saturated~\cite{awake}, at $z > 4$--$5 \; \mathrm{m}$, the potential for variations at $z < 4 \; \mathrm{m}$ to affect the final energy of the accelerated electrons is not critical.
More relevantly, after $z \approx 6\;\mathrm{m}$ all field values converge to that of the baseline case, within $\pm 2\%$.
This shows that the wakefield amplitude in these simulations is weakly dependent on the initial proton bunch parameters after 6~m along the plasma.

Before SSM saturation, i.e. where linear wakefield theory is still valid (before 4~m), the trends in Fig.~\ref{fig:e1averel} are consistent with $E_z\propto\frac{N_b}{\sigma_{zb}\sigma_{rb}^2}$: an increase of $N_b$ by +5\% produces higher values for $\left< |E_z| \right>$, for example, and the variations in $\sigma_{zb}$ and $\sigma_{rb}$ cause inversely proportional effects, with the $\sigma_{rb}$ parameter variations causing the largest effects.
There is also a clear effect on the growth rate, as evinced by the different slopes up to $z = 3 \; \mathrm{m}$ in Fig.~\ref{fig:e1ave}.

\begin{figure}
\begin{overpic}[width = \linewidth, tics=10]{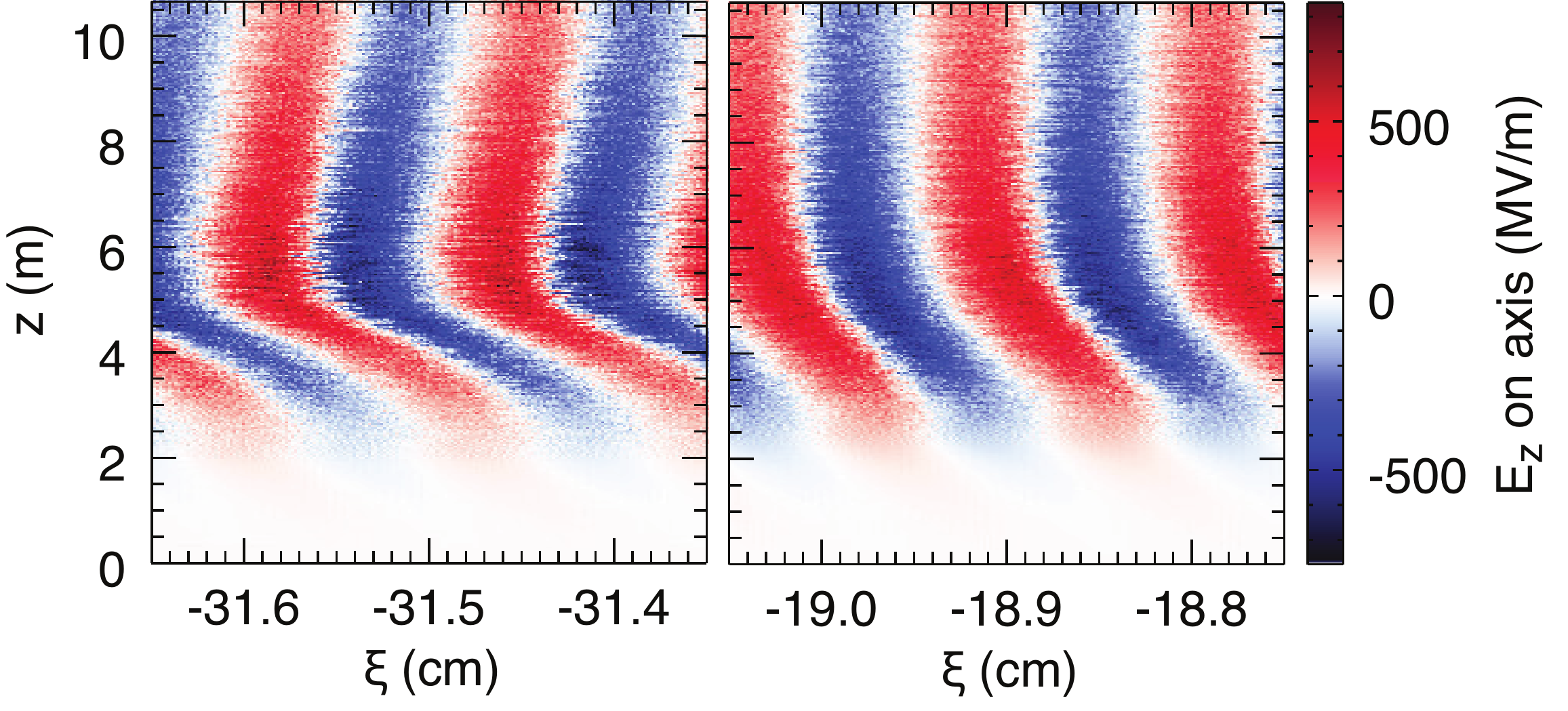}
	\put(12.5,13){\footnotesize (a)}
    \put(49.5,13){\footnotesize (b)}
\end{overpic}
\caption{\label{fig:e1vd} Waterfall plots of the line-outs of $E_z$ on the axis along the propagation distance $z$ for two different regions of the wakefields, (a)  $\xi \approx - 2.5 \: \sigma_{zb}$ and (b) $\xi \approx - 1.5 \: \sigma_{zb}$, where $\xi$ is defined with respect to the position of the seed.}
\end{figure}

Since the timing jitter $\Delta t$ is small when compared to the bunch duration we expect its main effect to be associated with an increase or decrease in total charge driving the wakefields (corresponding to $N_b$ variations by $\pm 2.85\%$).
With our choice of plasma radius ($r_p$), a $\pm$5\% variation seems to have no significant effect on the wakefield amplitude.
It has been shown that a smaller plasma radius can enhance the wakefield's focusing force and hence the SMI's growth rate by hindering the plasma's shielding response to the charge in the drive bunch~\cite{rp}.
However, this effect only becomes prominent when $r_p$ approaches $\sigma_{rb}$, which, despite the variations of $\pm 5 \%$, is not the case here.

We now turn our attention to the behavior of the wakefield phase.
Assuming that an electron is moving with a constant velocity $v_e$ in a region of the wakefields that is accelerating and focusing, when the phase velocity of the wakefields $v_\phi$ is below or above $v_e$, the electron will eventually slip out of this region and into an undesirable one (decelerating or defocusing).
This happens at the latest when the electron and the wakefields dephase by $\lambda_{pe}/4$ with respect to each other (in linear wakefield theory).

Numerical simulation results show that during the SSM growth the phase velocity of the wakefields varies along the plasma and along the bunch, eventually converging towards that of the driver after the SSM has saturated~\cite{schroeder,pukhov}.
This is also shown in Fig.~\ref{fig:e1vd}, where the evolution of the wakefield phase velocity is visualized by plotting the on-axis longitudinal field component $E_z$ in a waterfall plot along the plasma.
Since the simulation window moves at $c$, a negative slope in this type of graph means that the phase velocity of the wakefields is subluminal, while a positive slope indicates that it is superluminal.
The relativistic proton bunch moves at nearly the speed of light, so its velocity essentially corresponds to a vertical line in Fig.~\ref{fig:e1vd} (the slope $\frac{\Delta z}{\Delta\xi}\approx -2 \: \gamma^2$ for bunch particles).

We use the longitudinal component $E_z$ to characterize the evolution of the phase of the wakefields, as we did for the amplitude.
We make this choice because, though the transverse wakefields drive the SSM, they must be evaluated at the proper radius (e.g. at the bunch RMS transverse size for a Gaussian profile).
Since both transverse radius and shape of the bunch change as the SSM evolves, the evaluation becomes ambiguous.
In contrast, the longitudinal wakefield $E_z$ is well defined and maximum on the beam axis.
Moreover, the transverse and longitudinal wakefields share a fixed phase relationship due to the Panofsky-Wenzel theorem~\cite{PWtheo}, which means that the phase behavior can be measured through either component.

\begin{figure}
\begin{overpic}[width = \linewidth, tics=10]{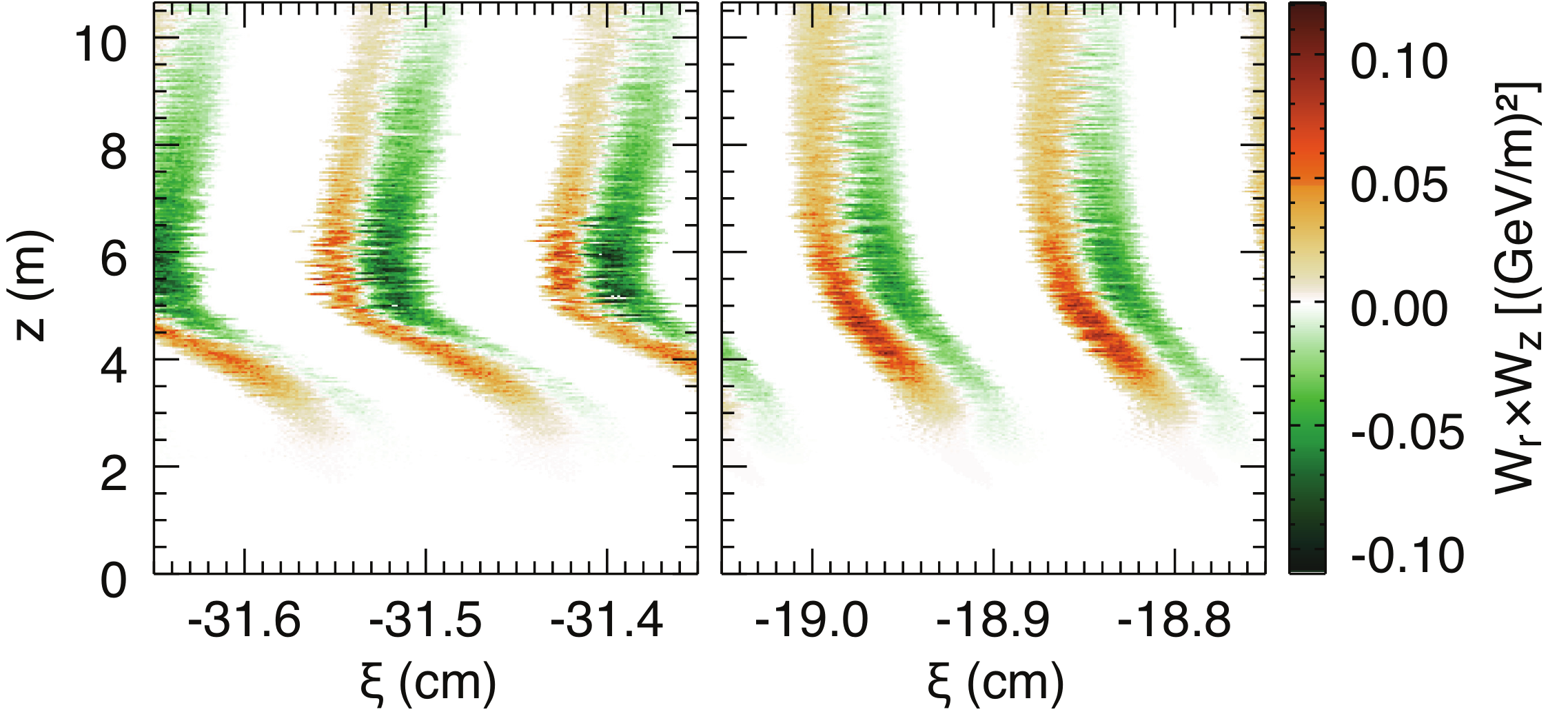}
	\put(12.5,13){\footnotesize (a)}
    \put(49.5,13){\footnotesize (b)}
\end{overpic}
\caption{\label{fig:wrwz} Waterfall plots of the product of the transverse and longitudinal forces $W_r$ and $W_z$ at $r = 0.7 \; k_{pe}^{-1}$ for (a) $\xi \approx - 2.5 \: \sigma_{zb}$ and (b) $\xi \approx - 1.5 \: \sigma_{zb}$, shown only for the accelerating regions ($W_z > 0$). The orange areas ($W_r > 0$) are defocusing for electrons and the green areas ($W_r < 0$) focusing.}
\end{figure}

To illustrate this last point, we produce a similar plot to the one in Fig.~\ref{fig:e1vd}, but for the product of the transverse and longitudinal force components $W_r$ and $W_z$, which, in cylindrical coordinates, are defined as $W_r = q \; (E_r - c \, B_\theta)$ and $W_z = q \; E_z$, respectively.
Here, $q$ is the charge of the test particle, $E_r$ is the radial component of the electric field and $B_\theta$ is the azimuthal component of the magnetic field.
The product $W_r \times W_z$ is evaluated at $r = 0.7 \; k_{pe}^{-1}$ in Fig.~\ref{fig:wrwz}, since the transverse components of the wakefields are zero on the axis, and we only consider the accelerating regions, i.e. where $W_z > 0$.

Figures~\ref{fig:e1vd} and~\ref{fig:wrwz} show that, while the wakefields are growing ($z < 4 \; \mathrm{m}$), they are slower than the drive beam velocity (negative slope).
In the region around $1.5\;\sigma_{zb}$ or 18.9~cm behind the seed [Figs.~\ref{fig:e1vd}(b) and~\ref{fig:wrwz}(b)] the phase velocity of the wakefields becomes essentially equal to the driver velocity after $z = 5 \; \mathrm{m}$ (vertical slope), which makes it a suitable position for the external injection of electrons.
Further behind the seed [around $-2.5 \; \sigma_{zb}$, Fig.~\ref{fig:e1vd}(a)] the phase velocity is superluminal for $z > 5 \; \mathrm{m}$, while earlier (for example around $\xi \approx - \sigma_{zb}$) it is subluminal (not shown).
Experimentally, the injection position along the bunch can be scanned so as to find the optimal $\xi$ position for maximum electron acceleration.

The effects of the parameter variations on the phase of the wakefields are studied quantitatively by fitting the function $f(\xi) = ~A\; \sin \left[ k_{pe} \left( \xi - \xi_s \right) + \phi \right]$ (expected for linear wakefields) to 2.5-$\lambda_{pe}$-long segments (starting at $\xi_s$) of the waterfall plots discussed in Fig.~\ref{fig:e1vd}, where $A$ and $\phi$ are the fitting parameters.
The value of $\phi$ is always relative to the seed position $\xi_s$.

\begin{figure}
\includegraphics[width = \linewidth]{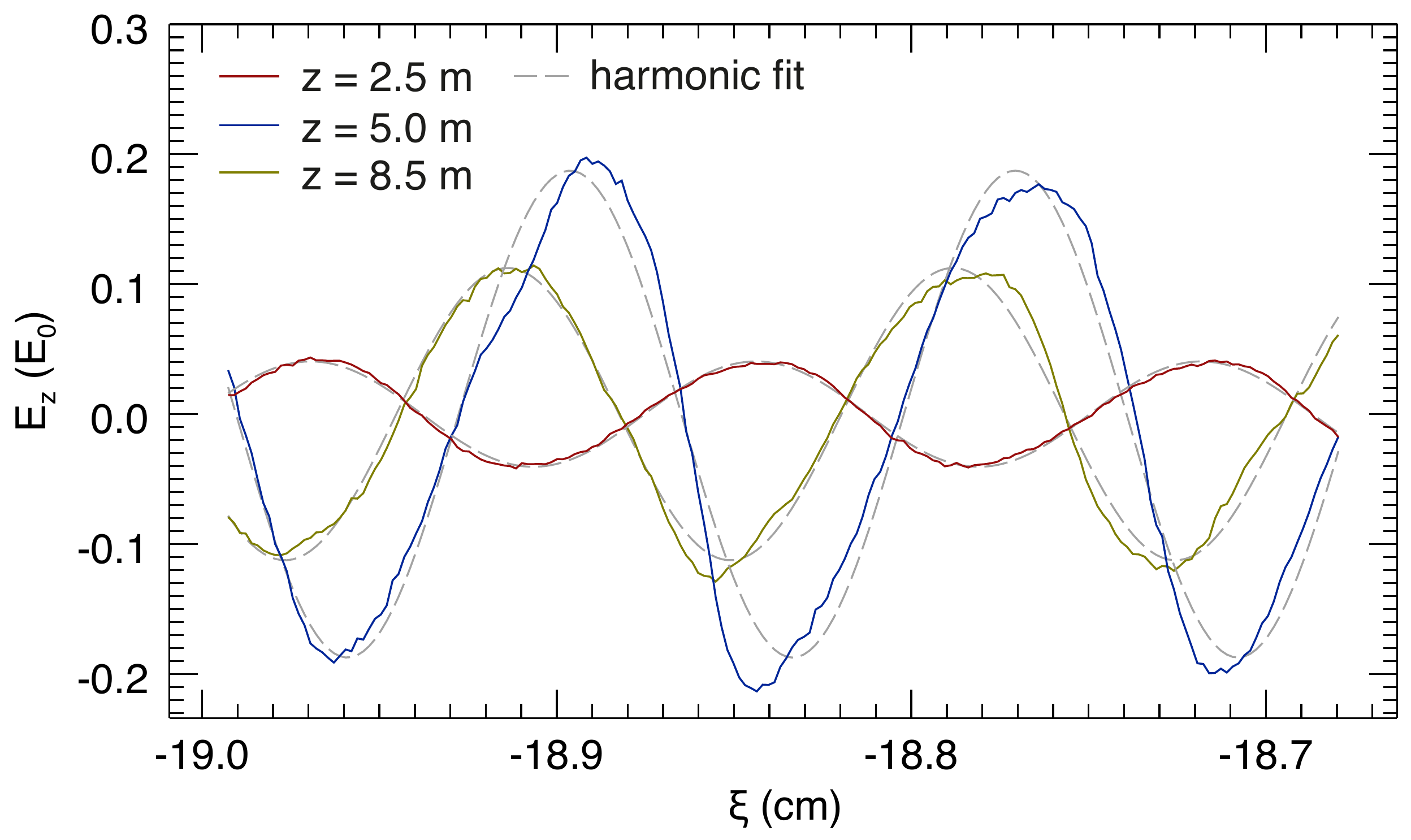}
\caption{\label{fig:fit}  Examples of the harmonic fit to a segment of $E_z$ located around $\xi \approx -1.5 \; \sigma_{zb}$ at three different positions along the plasma. The solid lines correspond to the data (smoothed) and the dashed lines to the fits.}
\end{figure}

As an example, the fit to a segment located around $\xi \approx - 1.5 \; \sigma_{zb}$ is shown in Fig.~\ref{fig:fit} for three different propagation distances. The fit is worst around the saturation point of the SSM (see curves for $z = 5 \; \mathrm{m}$), where the fields show signs of nonlinearity (the presence of high harmonics which lead to wave steepening).
However, the purpose of the fit is to define a local phase shift with respect to $\xi_s$, which is accomplished if the phases of both curves match, as is the case.

\begin{figure}[b]
\includegraphics[width = \linewidth]{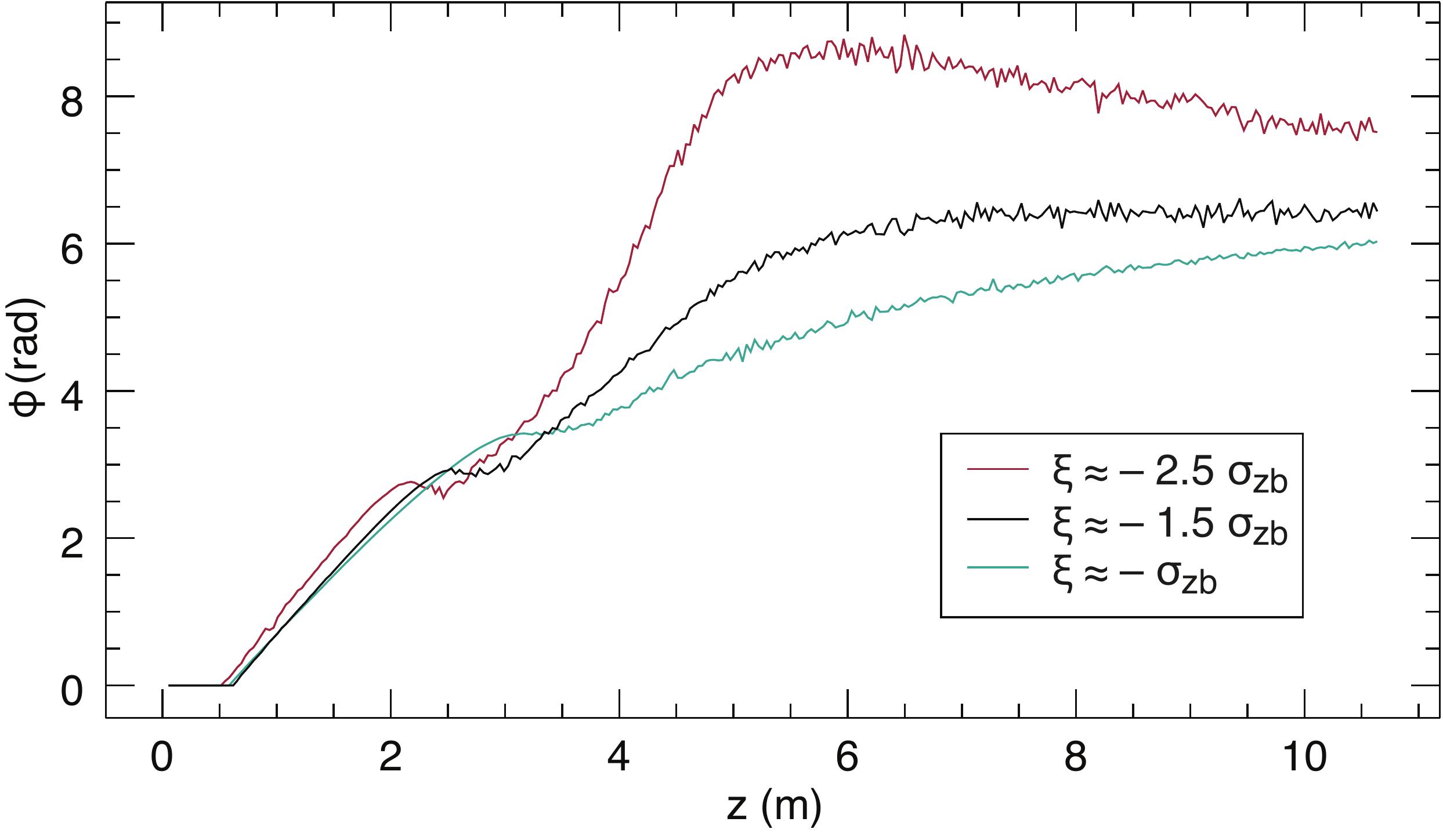}
\caption{\label{fig:phasepos} Evolution of the local phase shift of the wakefields with respect to the seed for three different positions along the proton bunch.}
\end{figure}

The result of this analysis for $\phi$ is shown in Fig.~\ref{fig:phasepos} for three different positions along the bunch.
Note that the burgundy and black curves correspond to the cases in Fig.~\ref{fig:e1vd}.
This figure again indicates that injection closer to $1.5 \; \sigma_{zb}$ rather than  $1.0 \; \sigma_{zb}$ behind the seed would be more beneficial, since a slower wakefield phase velocity leads to early dephasing.

\begin{figure}
\includegraphics[width = \linewidth]{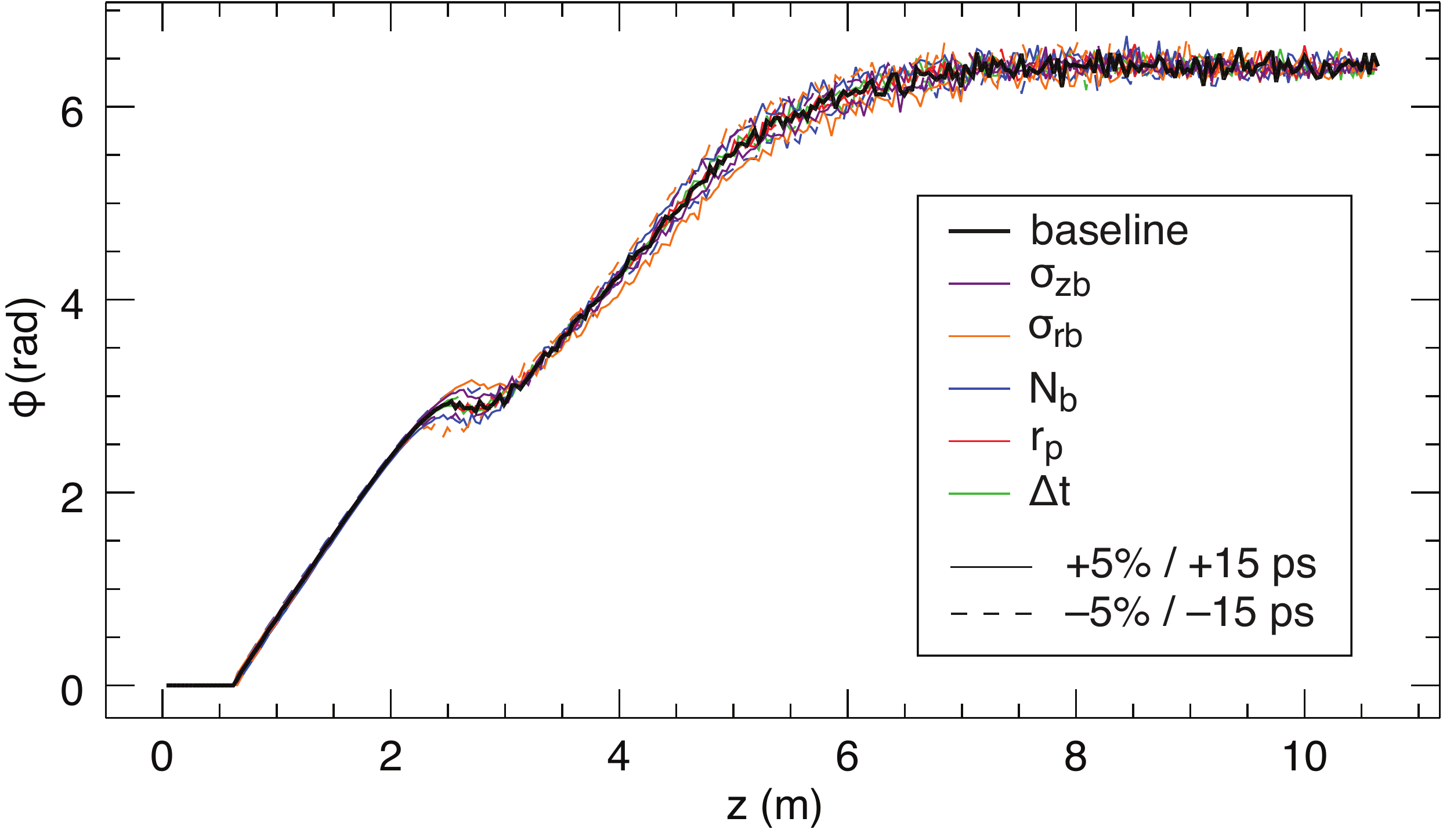}
\caption{\label{fig:phasecomp} Evolution of the local phase shift of $E_z$ with respect to the seed along the propagation distance $z$ at $\xi \approx - 1.5 \; \sigma_{zb}$ for different parameter variations.}
\end{figure}

The position $\xi \approx - 1.5 \; \sigma_{zb}$ was chosen for the comparison of the effects from the parameter scans, shown in  Fig.~\ref{fig:phasecomp}.
In the linear and strongly-coupled regime, i.e. before saturation and for $k_{pe} |\xi| \gg \frac{k_b}{\sqrt{2 \gamma_b}} z$ where $k_b = \sqrt{e^2 n_{b0} / \varepsilon_0 \, M_b} $ and $M_b$ is the mass of the drive bunch particles, the longitudinal wakefield component behaves approximately as $E_z \propto \cos \left[ k_{pe} \; \xi - \frac{\pi}{4} + \varphi(\xi,z) \right]$, with the phase shift $\varphi(\xi,z) \propto n_{b0}^{1/3}$~\cite{schroeder}.
The condition for the strongly-coupled regime is fulfilled for $\xi \approx - 1.5 \: \sigma_{zb}$ and $z \sim 10 \; \mathrm{m}$, with $k_{pe} |\xi| \approx 940.4$ and $\frac{k_b}{\sqrt{2 \gamma_b}} z \approx 1.9$.
Nevertheless, the phase shift in Fig.~\ref{fig:phasecomp} only displays a relationship of the form $\phi \propto \left( \frac{N_b}{\sigma_{zb}\sigma_{rb}^2 } \right)^{1/3}$ (after substituting Eq.~\ref{eq:nb0}) roughly between $z = 3.5$--$5\;\mathrm{m}$.

The largest effects on the wakefield phase are again observed before the saturation of the SSM, at $z = 2$--$3 \; \mathrm{m}$ (see Fig.~\ref{fig:phasecomp}).
Here, the largest difference is of roughly $2 \pi/20$ for $0.95 \; \sigma_{rb}$ at $z \approx 2.5\;\mathrm{m}$.
After this point, phase variations are limited to $\pm$0.4~rad (corresponding to approximately $\lambda_{pe}/16$), an estimate constrained by simulation noise.
Moreover, the phase stops changing after $z \approx 6\;\mathrm{m}$ in all cases, which is also the point after which the wakefield amplitude becomes essentially independent of the proton bunch parameter variations (see Fig.~\ref{fig:e1averel}).

This suggests that, at this plasma density and for the chosen proton bunch parameters, electrons injected at $z \approx 6 \;\mathrm{m}$ or further remain in phase with the wakefields for a long distance and can therefore be accelerated to high energies in wakefields with a constant phase.

\section{Behavior of accelerated electrons}

AWAKE aims to demonstrate the acceleration of an electron bunch, and therefore it is important to study the effects of initial parameter fluctuations on the properties of these electrons and not only on the wakefields, as was done so far.
The characteristics of the accelerated electron bunch are the most important experimental output, and they are non-trivially dependent on several factors besides the wakefields themselves, such as the electrons' initial velocity or the injection point along the plasma.
Consequently, the wakefield variations reported above are not sufficient to infer possible effects on the accelerated bunches.

A simple diagnostic was devised to determine the energy gain acquired by an electron as a function of its injection point along the plasma ($z_\mathrm{inj})$ and its initial position along the bunch ($\xi_0$).
This algorithm is in practice a one-dimensional (1D) particle pusher: for each possible $z_\mathrm{inj}$ along the plasma, a test particle is placed at $\xi_0$ along the on-axis wakefield (i.e. the data presented in Fig.~\ref{fig:e1vd}) and propagated forwards in the wakefields.
All test electrons have an initial energy corresponding to $\gamma_0 = 39.1$, or approximately 20~MeV (the maximum range of the electron injector commissioned for AWAKE~\cite{awake}).

\begin{figure}
\begin{overpic}[width = \linewidth,tics=10]{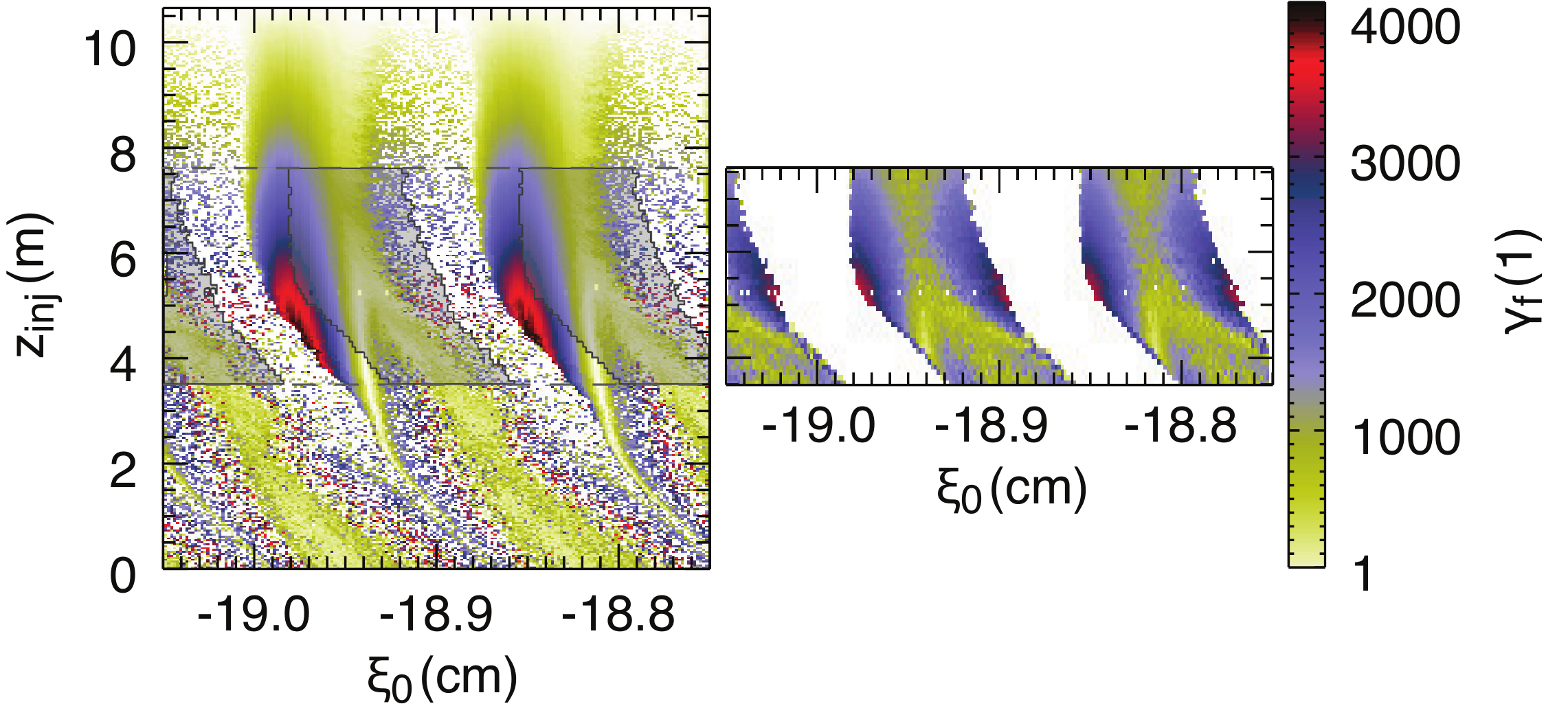}
	\put(39,41){\footnotesize (a)}
    \put(76,38){\footnotesize (b)}
\end{overpic}
\caption{\label{fig:energy} (a) Final energy of on-axis test electrons that reach the end of the plasma ($z = 10\;\mathrm{m}$) according to their injection point $z_\mathrm{inj}$ and their initial position along the bunch $\xi_0$, as obtained from the 1D pusher, and (b) peak final energy of electrons initially fulfilling $r_0 < 0.5\; k_{pe}^{-1}$ that reach the end of the plasma for several injection points and initial positions, as obtained from a PIC simulation. White areas mean no electrons. The shapes in (b) are superimposed in (a) as shadowed areas, for comparison.}
\end{figure}

The spatial resolution of these results is limited to the resolution of the simulation box in the $\xi$ direction (which in this case means that at most 38 evenly-spaced test electrons can be tracked for every $\lambda_{pe}/2$), while the temporal resolution is limited to the number of simulation file dumps (in this case 300 over 10~m, giving a maximum resolution for $z_\mathrm{inj}$ of 3.55~cm).
In this diagnostic, the electrons are assumed to remain on the axis at all times and no transverse forces are considered.
Tracking particles in axisymmetric two-dimensional space (including transverse fields) would in effect entail full-fledged PIC simulations.

Since $E_z$ peaks on the axis and decays radially, an electron performing any transverse motion about the axis is subject to weaker longitudinal forces than if it is propagating exclusively along it (the most effective trajectory in terms of energy gain).
This approach thus provides a best case scenario for the energy gained by accelerated electrons.
It nonetheless includes their dephasing with respect to the wakefields, while the simplicity of the approach means that results can be obtained quickly for many different cases, e.g. for different injection points and for all the parameter scans performed in this work.

The result of this diagnostic is shown in Fig.~\ref{fig:energy}(a) for the baseline simulation as a scatter plot of electrons that reach the end of the plasma, with their energy (color-coded) as a function of their injection position ($\xi_0$,$z_\mathrm{inj}$).
The rest of the test electrons lose enough energy at some point along $z$ so as to slip out of the 33-centimeter-long simulation window, and hence not reach the end of the plasma.

The general features of the accelerating field [see Fig.~\ref{fig:e1vd}(b)] are visible in the point density of Fig.~\ref{fig:energy}(a).
Regions with few test electrons correspond to decelerating regions.
In regions where the field is accelerating ($E_z < 0$, for example $-19.00 < \xi_0 \: [\mathrm{cm}] < -18.95$), all the test electrons reach the end of the plasma.
As expected, the final energies decrease as electrons are injected at later $z$ positions (shorter acceleration distances), though this is also because the wakefield amplitude decreases after $z \approx 5\; \mathrm{m}$ (see Fig.~\ref{fig:e1ave}).
Figure~\ref{fig:energy}(a) also implies that some electrons injected in the decelerating phase of the wakefields survive energy loss and dephasing to ultimately reach large energies (scattered red dots). The same is true for electrons injected before the saturation of the SSM ($z < 4 \; \mathrm{m}$).

\begin{figure}
\includegraphics[width = 0.57\linewidth,tics=10]{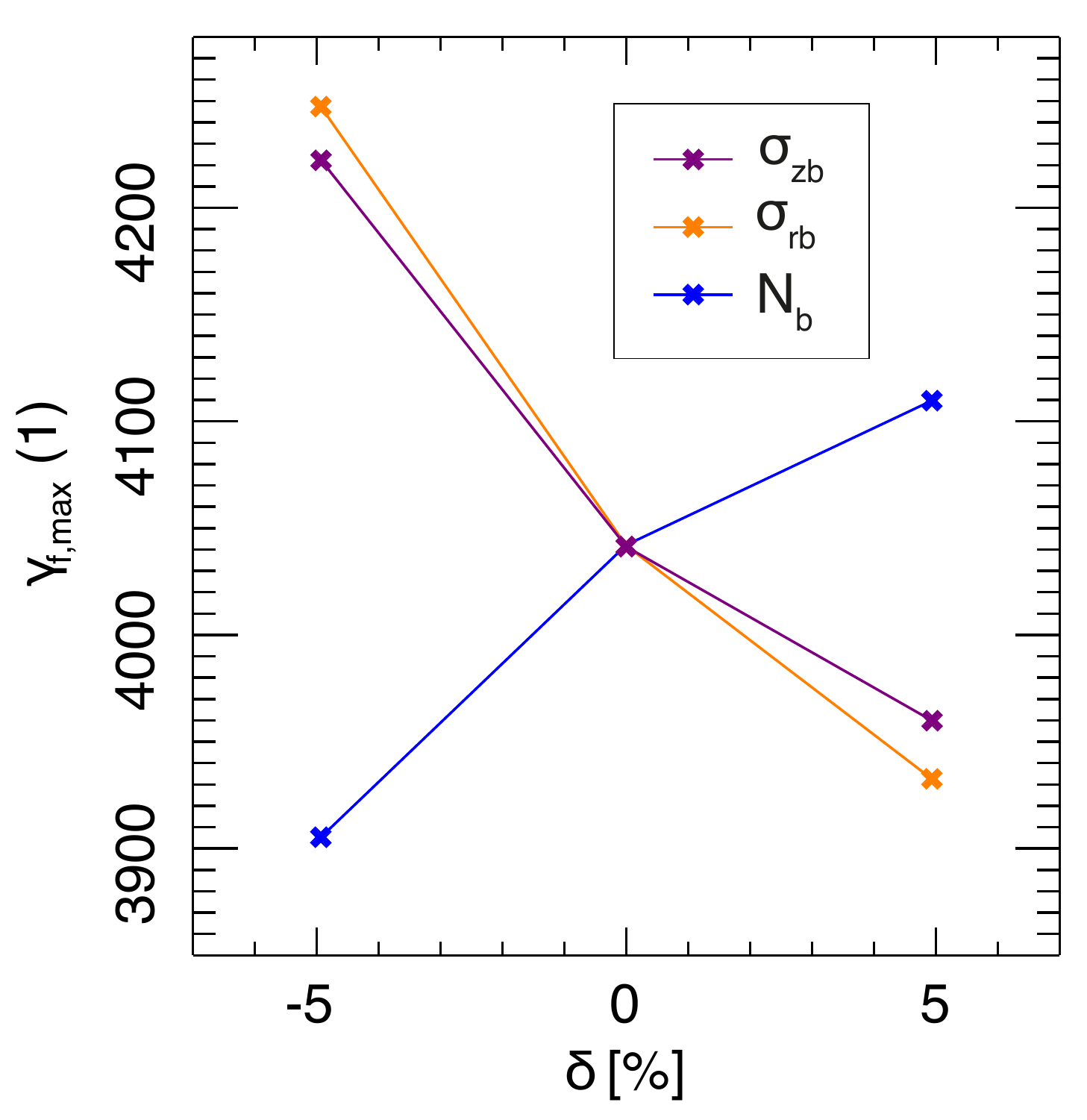}
\caption{\label{fig:emaxpars} Maximum final energy $\gamma_{f,\mathrm{max}}$ found through the 1D diagnostic of Fig.~\ref{fig:energy}(a) for three different parameter variations ($\delta = 0$ corresponds to the baseline parameters).}
\end{figure}

The diagnostic was applied to the bunch parameter scans to evaluate possible effects on the final energy of injected electrons.
We compared the maximum final energy attained by test electrons injected in the same wakefield period for each parameter variation, choosing the range $ -18.990 \le \xi_0 \: [\mathrm{cm}] \le  -18.956 $ (approximately $\lambda_{pe}/4$-long), where $\gamma_f$ is maximal.

Figure~\ref{fig:emaxpars} shows the scatter plot of $\gamma_{f,\mathrm{max}}$ for the variations $\delta$ of $\sigma_{zb}$, $\sigma_{rb}$ and $N_b$, the parameters that caused the largest effects.
We find trends of the form $\gamma_{f,\mathrm{max}} \propto\frac{N_b}{\sigma_{zb}\sigma_{rb}^2}$, which is consistent with the behavior observed above for the average wakefield amplitude $\left< |E_z| \right>$ and with the fact that the energy gain by trailing particles is directly linked to the amplitude of the axial field component $E_z$.
The resulting maximum final energies vary at most between roughly $-3\%$ and $+5\%$ (the corresponding injection points lie between 4.15 and 4.52~m along the plasma).

To validate the diagnostic described above, we performed a full simulation with the baseline parameters, in which test electrons were injected at 41 equally-spaced injection points between 3.5~and 7.6~m.
The electrons used in the simulation have zero emittance and are initially uniformly distributed in space (both longitudinally and transversely).
The electron data was processed so as to obtain the same type of graph as Fig.~\ref{fig:energy}(a).
This data is shown in Fig.~\ref{fig:energy}(b) for electrons injected close to the axis ($r_0 < 0.5 \; k_{pe}^{-1}$) that reached the end of the plasma.

We would expect to observe the influence of the transverse wakefields in the final energy distribution on the $(\xi_0,z_\mathrm{inj})$ plane of Fig.~\ref{fig:energy}(b), which is indeed the case.
The regions of electron loss in Fig.~\ref{fig:energy}(b) (due to transverse forces) are much clearer than those on Fig.~\ref{fig:energy}(a) (which are only due  to longitudinal dephasing).
The periodic regions with the most electrons in both plots [i.e. accelerating phases in (a) and focusing phases in (b)] also appear to be shifted by around $\lambda_{pe}/4$ with respect to each other [note the shape of the scatter plot in (b) superimposed on (a)], as would be expected from the Panofsky-Wenzel theorem~\cite{PWtheo}.
Other than this, the overall distribution matches well with that of Fig.~\ref{fig:energy}(a).

\begin{figure}
\includegraphics[width = \linewidth]{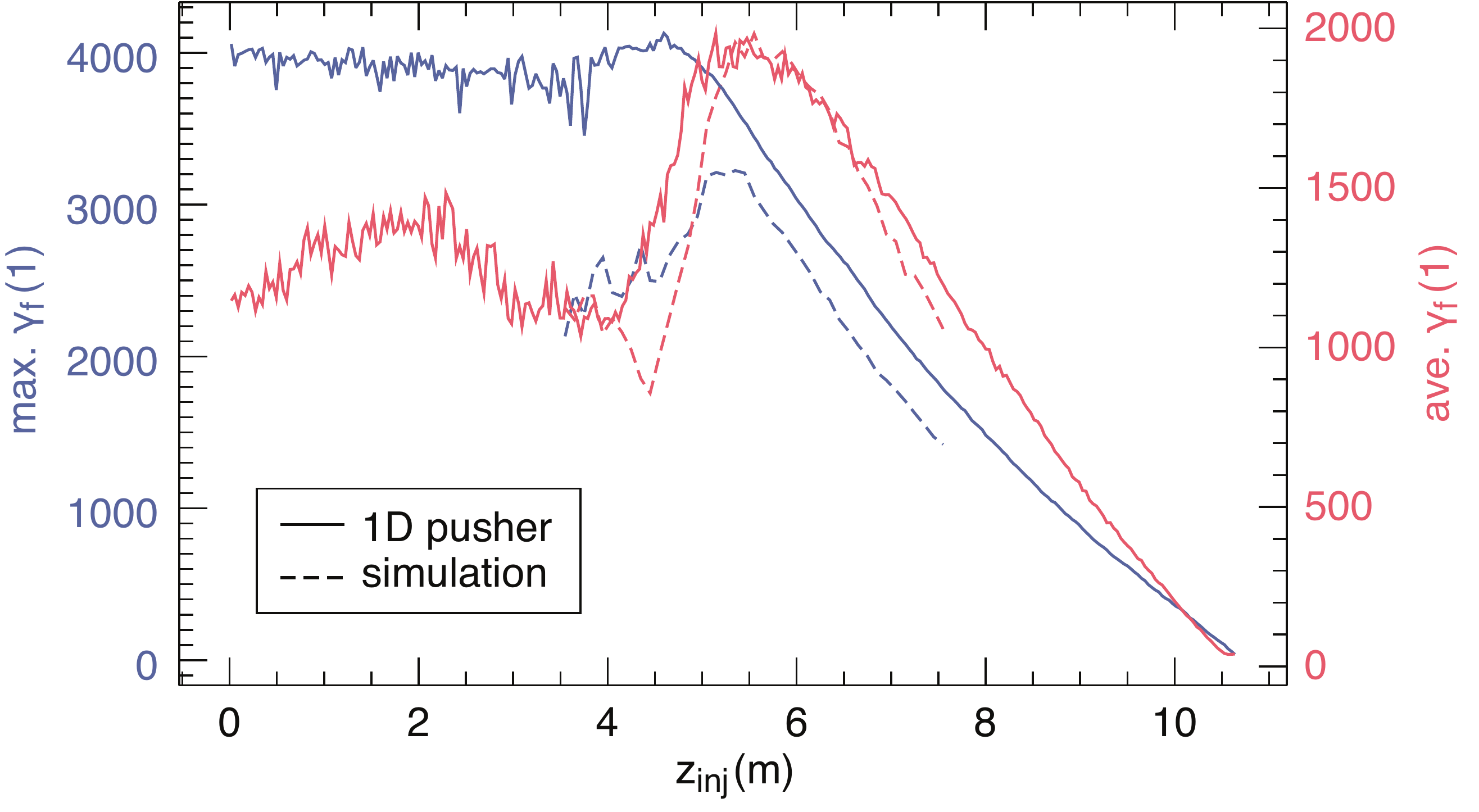}
\caption{\label{fig:gfmaxave} Maximum (blue) and average (pink) final energies for each row of the 1D diagnostic graphs in Fig.~\ref{fig:energy}(a) (solid lines) and PIC simulation results \ref{fig:energy}(b) (dashed lines) as a function of the injection point $z_\mathrm{inj}$.}
\end{figure}

A more quantitative comparison of the 1D pusher with direct simulation results can be seen in Fig.~\ref{fig:gfmaxave}, which shows the average along each row of both graphs in Fig.~\ref{fig:energy} (pink) as well as each row's maximum energy (blue) plotted against the injection point $z_\mathrm{inj}$.

The peak energies in the 2D simulation results are generally lower than the 1D results (compare dashed and solid blue curves), which is expected since the 1D diagnostic represents a best-case scenario.
Furthermore, their trends do not agree before $z_\mathrm{inj} = 5.5\;\mathrm{m}$.
This is the region where we expect the variation of the wakefield phase and associated defocusing to be the largest.
For $z_\mathrm{inj} > 5\;\mathrm{m}$, however, where we expect these effects to be negligible, the trend in both curves is very similar.
The average energies in turn show very good agreement (pink curves).

We can therefore conclude that the 1D diagnostic was an appropriate tool for a comparative analysis of the effects of the parameter variations on the final energies of electrons that are initially close to the axis.

\begin{figure}
\begin{overpic}[width = \linewidth,tics=10]{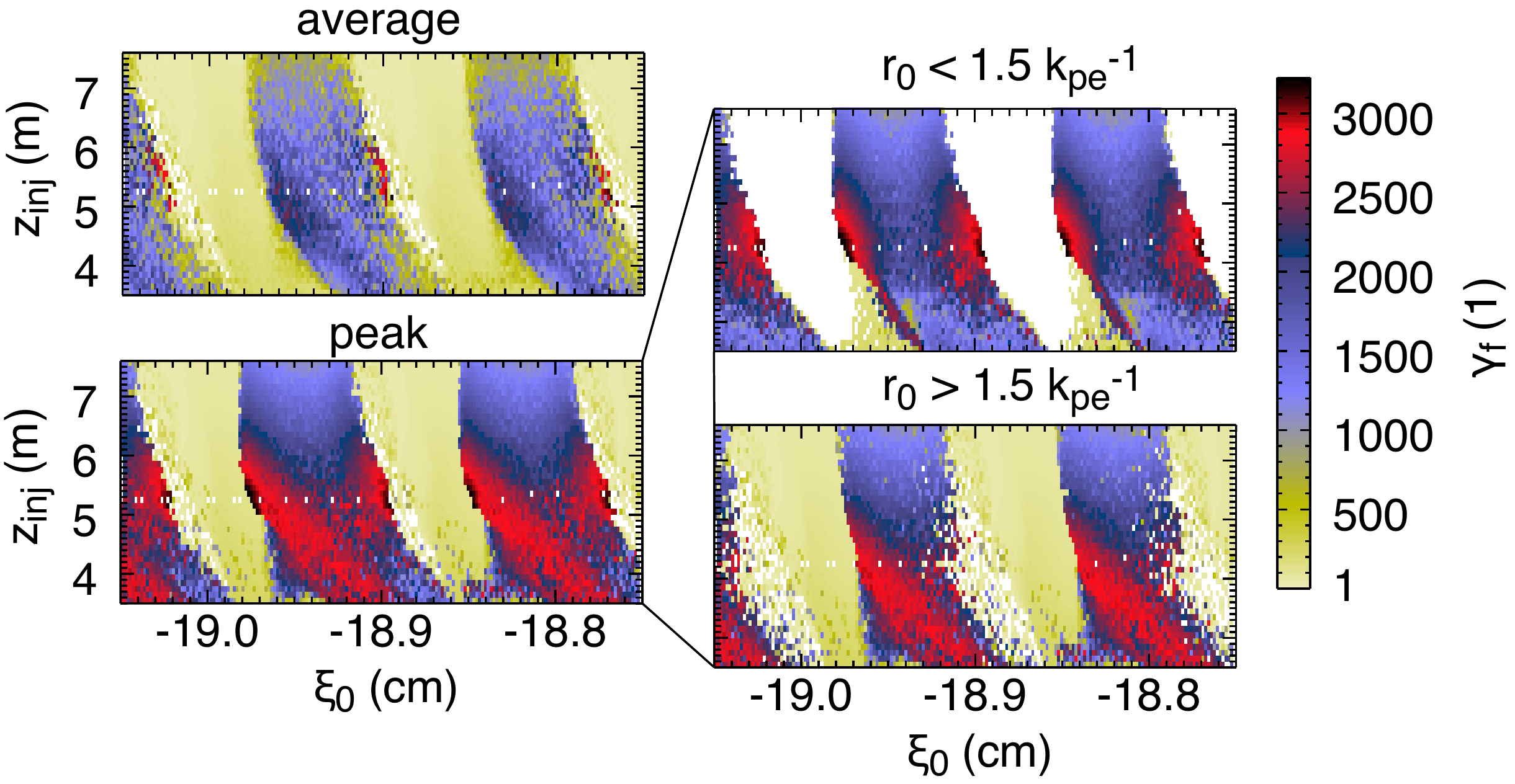}
	\put(10,44){\footnotesize (a)}
    \put(10,23.5){\footnotesize (b)}
    \put(49,40.5){\footnotesize (c)}
    \put(49,19.5){\footnotesize (d)}
\end{overpic}
\caption{\label{fig:gfsim} (a) Radial average and (b)--(d) peak final energies of injected electrons as a function of their initial longitudinal position $\xi_0$ and their injection point $z_\mathrm{inj}$, as obtained from the PIC simulation. The peak energies are shown for (b) all initial radii $r_0$, (c) for $r_0 < 1.5\;k_{pe}^{-1}$, and (d) for $r_0 > 1.5\;k_{pe}^{-1}$. }
\end{figure}
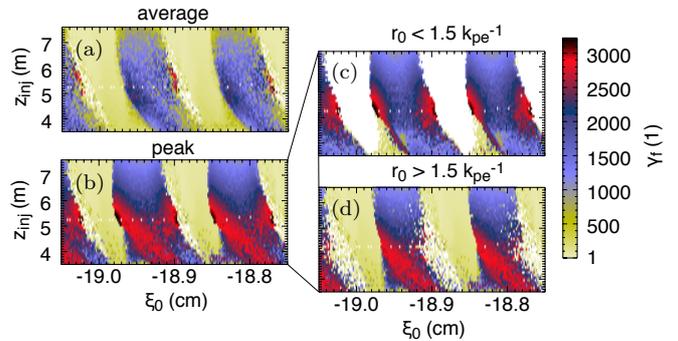

The peak energy curve obtained from the simulation in Fig.~\ref{fig:gfmaxave} (dashed blue line) suggests that the optimum injection point lies between 5--6~m.
Although this graph only represents electrons initially close to the axis ($r_0 < 0.5 \; k_{pe}^{-1}$), the optimal injection range is confirmed when the final energies of all electrons at all possible radii (up to the plasma boundary $r_p$) are considered, as shown in  Figs.~\ref{fig:gfsim}(a) and (b).
Each data point in Fig.~\ref{fig:gfsim}(a) consists of an average of all the simulation particles that began at a given $\xi_0$ over the entire plasma radius, while Fig.~\ref{fig:gfsim}(b) shows the peak energy out of all electrons with any $r_0$ for each $\xi_0$.
Both scatter plots display the highest Lorentz factors for $z_\mathrm{inj} = 5$--$6\;\mathrm{m}$.

Figure~\ref{fig:gfsim}(b) furthermore indicates that some electrons reach high energies when injected before saturation of the SSM ($z_\mathrm{inj} < 5\;\mathrm{m}$), which is also suggested by the 1D diagnostic results [note red points for $z_\mathrm{inj} = 0$--$5\;\mathrm{m}$ in Fig.~\ref{fig:energy}(a)].
When we decompose the data in Fig.~\ref{fig:gfsim}(b) into electrons originating above and below a radius of $1.5\;k_{pe}^{-1}$ (approximately 0.3~mm), we find that, for injections before 5~m, the electrons far from the axis attain the highest energies [Fig.~\ref{fig:gfsim}(d)], while for $z_\mathrm{inj} = 5$--$6\;\mathrm{m}$ it is the electrons close to the axis that gain the most energy [Fig.~\ref{fig:gfsim}(c)].

\begin{figure}[b]
\begin{overpic}[width = \linewidth,tics=10]{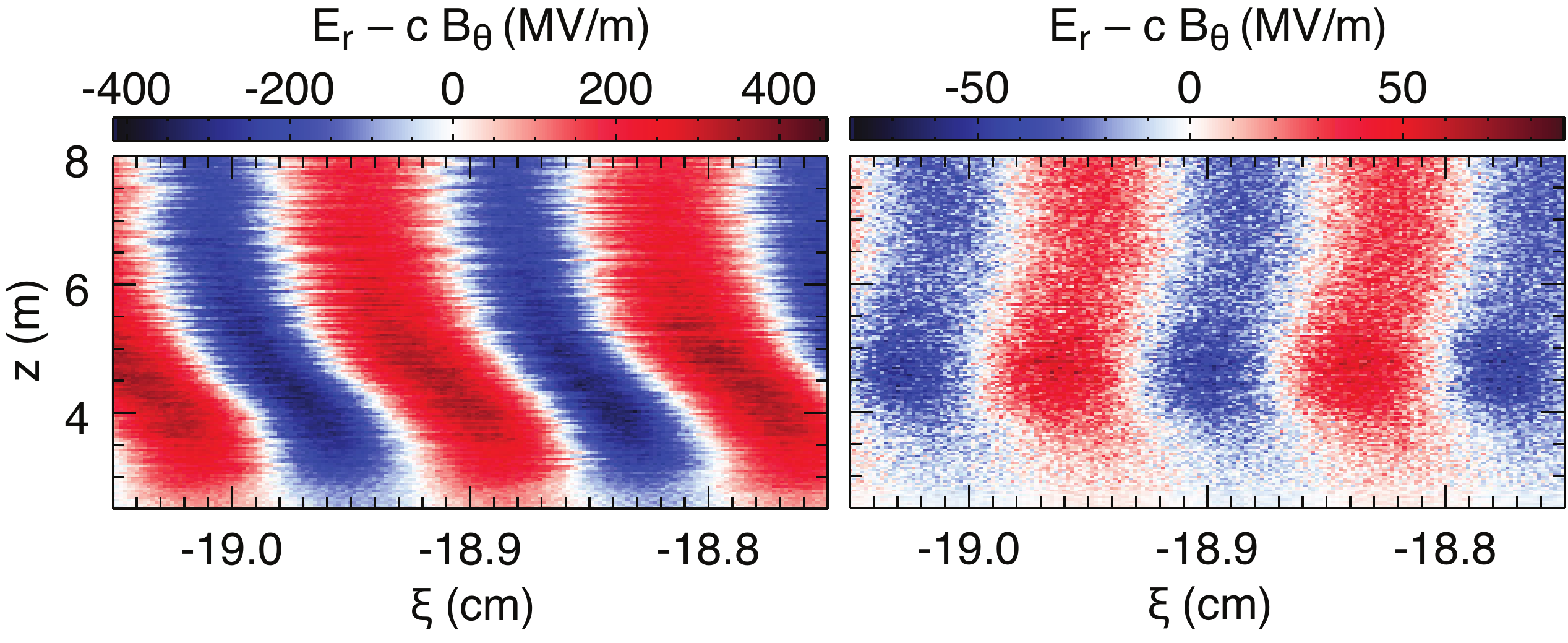}
	\put(7,38.5){\footnotesize (a)}
    \put(54,38.5){\footnotesize (b)}
\end{overpic}
\caption{\label{fig:e2b3} Waterfall plots of $E_r - c \, B_\theta$ for $\xi \approx - 1.5 \; \sigma_{zb}$ at (a) $r = k_{pe}^{-1}$ and (b) $r = 3 \; k_{pe}^{-1}$. }
\end{figure}

This difference is only observable for injections that take place before saturation of the SSM and could thus be explained by its development.
In fact, the PIC simulations show that the phase velocity of the wakefields varies along the plasma radius as well.
This is demonstrated by the waterfall plots of the transverse wakefield component $E_r - c \, B_\theta$ (which is responsible for focusing and defocusing) in Fig.~\ref{fig:e2b3}.
For $z < 5 \; \mathrm{m}$, for example, the phase velocity closer to the axis [Fig.~\ref{fig:e2b3}(a), at $r = k_{pe}^{-1}$] behaves as expected during the growth of the SSM and as previously discussed in Fig.~\ref{fig:e1vd}.
At a larger radius, however, the phase is approximately stable between 4~and 5.5~m [Fig.~\ref{fig:e2b3}(b), at $r=3\;k_{pe}^{-1}$].
This would explain why electrons starting before $z = 5\;\mathrm{m}$ at smaller radii would tend to be lost (due to the rapidly changing phase and their subsequent slippage into defocusing half-periods), while electrons further away from the axis would find a stable wakefield phase and thus gain energy over a larger distance.

\section{Summary}

Using PIC simulations, we varied the bunch parameters $\sigma_{zb}$, $\sigma_{rb}$ and $N_b$, the plasma radius $r_p$, and the seed point timing $\Delta t$, and studied their effect on the wakefield amplitude and phase during the development of a seeded instability (SSM), and on the maximum energy gain as determined by test electrons.

We found that the parameter variations we considered ($\pm$5\% and $\pm$15~ps) essentially lead to differences in wakefield amplitude and phase only in the growth region of the SSM along the plasma ($z<4\;\mathrm{m}$ in this case).
The wakefield parameters all converge to similar values after saturation of the SSM, within a few percents for the amplitude and the equivalent of less than $\lambda_{pe}$/8 for the phase.
While the results presented here were obtained for only one set of baseline parameters, the same analysis with different parameters showed similar trends.
Furthermore, it is clear that in practice all initial parameters vary for each event.
However, as variations may have counteracting effects, we assume that the conclusions reached through single parameter variation studies are still representative of experimental situations.

Based on the simulations, we also found that the optimal injection coordinates for our parameters ($n_0 = 7 \times 10^{14} \; \mathrm{cm^{-3}}$ and $N_b = 1.5 \times 10^{11}$) are 5--6~m into the plasma and around $1.5 \; \sigma_{zb}$ behind the wakefield seed. For an injection in this range, electrons close to the axis can reach energies of the order of 1.6~GeV over the last 4--5~m of plasma.
Comparable final energies are also attained when injection takes place before saturation of the SSM ($z < 5\;\mathrm{m}$), but by electrons far from the axis instead.

In general, the optimal injection point along the plasma will be determined by the start of the saturation of the SSM, which takes place earlier with either larger $n_0$ or $N_b$. The position with the most stable phase along the bunch can also be scanned for different parameters, and it tends to be closer to the seed point for higher $n_0$ and smaller $N_b$. The increase of either of these two parameters will further lead to higher wakefield amplitudes, and hence to larger energy gains by trailing electrons.
In the future, we will seek further optimization towards a higher accelerated beam quality, for example by including the witness beam emittance and beam loading effects in PIC simulations of the entire injection and acceleration process (see for example~\cite{olsen}).

\begin{acknowledgments}

M. M. acknowledges the scientific guidance provided by Bernhard Holzer.
We acknowledge the grant of computing time by the Leibniz Research Center on SuperMUC.
J. V. acknowledges the support of FCT (Portugal) Grant No. SFRH/IF/01635/2015 and CERN/FIS-TEC/0032/2017.

\end{acknowledgments}

\bibliography{refs}

\end{document}